\documentclass[12pt,letterpaper]{article}
\usepackage{amsthm}
\usepackage[margin=1in]{geometry}
\usepackage{algorithm}
\usepackage{bm}
\usepackage{color}
\usepackage{setspace}
\usepackage{caption2}
\newcommand{\blue}[1]{{\textcolor{black}{#1}}}

\renewcommand{\bar}{\overline}
\usepackage{booktabs}
\usepackage{amsfonts,amssymb,graphics,epsfig,verbatim,bm}
\usepackage{latexsym,amsmath,url,amsbsy}
\usepackage{multirow}
\usepackage{natbib}
\usepackage{xr-hyper}
\usepackage[colorlinks,citecolor=blue,urlcolor=blue,linkcolor=blue,
filecolor=blue,backref=page]{hyperref}
\usepackage{graphicx}
\usepackage{physics,bm}
\usepackage{algpseudocode}
\usepackage[mathscr]{euscript}
\numberwithin{equation}{section}
\theoremstyle{plain}

\newcommand{\TB}{\mbox{Bin}}
\renewcommand{\hat}{\widehat}
\renewcommand{\bar}{\overline}
\def\m{\mathcal}
\newtheorem{thm}{Theorem}[section]
\newtheorem{corollary}{Corollary}

\usepackage{todonotes}

\title{Bayesian Spatial Homogeneity Pursuit of 
Functional Data: an Application to the U.S.  Income Distribution}

\author{Guanyu Hu~~~~Junxian Geng~~~~Yishu Xue~~~~Huiyan Sang}
\begin{document}

\maketitle

\begin{abstract}
An income distribution describes how an entity's total wealth is distributed
amongst its population.  A problem of interest to regional economics
researchers
is to understand the spatial homogeneity of income distributions among
different
regions. In economics, the Lorenz curve is a well-known functional
representation of income distribution. In this article, we propose a mixture of
finite mixtures (MFM) model as well as a Markov random field constrained
mixture
of finite mixtures (MRFC-MFM) model in the context of spatial functional data
analysis to capture spatial homogeneity of Lorenz curves. We design efficient
Markov chain Monte Carlo (MCMC) algorithms to simultaneously infer the
posterior
distributions of the number of clusters and the clustering configuration of
spatial functional data. Extensive simulation studies are carried out to show
the effectiveness of the proposed methods compared with existing methods.  
We apply the proposed spatial functional clustering method to state level
income
Lorenz curves from the American Community Survey Public Use Microdata Sample
(PUMS) data. The results reveal a number of important clustering patterns of
state-level income distributions across US. 
\bigskip 

\end{abstract}

\section{Introduction}\label{sec:intro}

Our study is motivated by an American Community Survey Public Use Microdata
Sample (PUMS) data that contains incomes of United States (US) households in
year 2017, which can be accessed via the PUMS data registry
(\url{https://www.census.gov/programs-surveys/acs/data/pums.html}). Incomes of
households as well as the states they live in are recorded. Our primary goal is
to cluster the state level Income Distribution
\citep[ID;][]{sullivan2003economics}, i.e., how a state's total wealth is
distributed amongst its population. In order to clarify the differences between
economics term ``Income Distribution'' and density distribution of household
income, we use ID to represent this particular economic term in the rest of the
paper. The ID has been a central concern of economic theory since the time of
classical economists such as Adam Smith and David Ricardo. While economists have
been conventionally concerned with the relationship between factors of
production, land, labor, and capita for ID, modern economists now focus more on
income inequality. Particularly, a balance between income inequality and
economic growth is a desired goal for policy makers. Capturing homogeneity
pattern of state level IDs is of great research interest in economic studies, as
it will enhance the understanding of income inequality among different regions
within a country, and provide policy makers with reference as to issue different
policies for the identified regions. In macroeconomics, most governments want to
obtain an equitable (fair) distribution of income, which is a crucial element
of a functioning democratic society \citep{mankiw2014principles}. In order to
obtain this goal, the distribution of income or wealth in an economy is
represented by a Lorenz curve \citep{lorenz1905methods}, which is a function
showing the proportion of total income assumed by the bottom $100p\%$ ($p\in
[0,1]$) of the population. Derived from the Lorenz curve, the Gini coefficient
is a commonly used measure for income inequality \citep{gini1997concentration},
and it has been widely adopted by many international organizations, such as the
United Nations and World Bank, to study income inequalities among regions. The
Gini coefficient, however, is only a summary measurement of statistical
dispersion of ID, and it is non-unique as two Lorenz curves can assume different
shapes but still yield the same Gini value. \blue{The Gini index is related to
the Lorenz curve as twice the area between the 45-degree line and the Lorenz
curve which is insensitive to the changes of the form of the Lorenz curve.}
Similarly, the Hoover index \citep{hoover1936measurement} is also derived from
the
Lorenz curve, and suffers from the same non-uniqueness disadvantage.

Thus far, many methods have been introduced to either directly model Lorenz
curves or indirectly through the modeling of statistical distribution
functions of household income. Popular
parametric methods for \blue{modeling the density of personal incomes} in
general use heavy tail
distributions, including Pareto \citep{pareto1964cours},  log-normal
\citep{gibrat1931inegalites}, Weibull \citep{bartels1975alternative}, gamma
\citep{bartels1975alternative}, and generalized beta distributions
\citep{mcdonald1984some,mcdonald1995generalization}. Nonparametric methods
include the commonly used empirical Lorenz curve estimation method and several
other extensions that introduce various smoothing techniques \citep{ryu1996two,
cowell2008modelling}. Most of these existing methods only focus on modeling a
univariate personal ID. There is a need for the development of
spatial  functional data analysis techniques to jointly model Lorenz curves
across counties or states in economic studies. Without spatial homogeneity
pattern detection,  each state needs to make its own policy, which could
be a waste of public resource, while with a few
clusters of states, only (the number of clusters) policies need to be made
accordingly.

There are several major challenges in developing clustering algorithms for
spatial functional data. First, spatial functional data such as state-level
Lorenz curves often exhibit strong location-related patterns. It is necessary
to
incorporate such spatial structure into spatial functional data clustering
algorithms. Nevertheless, most existing functional clustering algorithms are
designed under the assumption that the observed functions are i.i.d curves
\citep[e.g., see a review paper by][]{jacques2014functional}. These methods can
be broadly classified into three paths: two-stage methods that reduce the
dimension by basis representations before applying clustering approaches,
nonparametric methods that define specific dissimilarities among functions
followed by heuristics or geometric procedures-based clustering algorithms such
as~$K$-means, and model-based methods that specify clustering models such as
mixture of Gaussian for basis coefficients. Recently, a number of works have
been proposed to extend these functional clustering algorithms to the spatial
context. \cite{romano2011clustering} and \cite{giraldo2012hierarchical}
followed
the second path to define dissimilarities among spatial functions based on
spatial variograms and cross-variograms. \cite{jiang2012clustering} followed
the
third path to model cluster memberships using an auto-regressive Markov random
field, and introduce spatially dependent random errors in the conditional model
for functions.

Second, it is desired to impose certain spatial contiguous constraints on the
clustering configuration to facilitate interpretations in the spatial context.
In other words, a local cluster is expected to contain spatially connected
components with flexible shapes and sizes. In addition, in many economics
applications, this spatial contiguous constraint may not dominate the
clustering
configuration globally, in the sense that two clusters that are spatially
disconnected may still belong to the same cluster. For example, New England
area
could share certain similar demographic information with California despite the
distance in between. Although a large body of model based spatial clustering
approaches have been proposed in various spatial contexts, to the best of our
knowledge, there is still a lack of clustering methods that allow for both
locally spatially contiguous clusters and globally discontiguous clusters. For
example, existing Bayesian spatial clustering methods based on mixture models,
such as the finite mixture model used in the aforementioned spatial functional
clustering algorithm \citep{jiang2012clustering}, can introduce spatial
dependence in cluster memberships but may not fully guarantee spatial
contiguity. \citet{suarez2016bayesian} clustered each
signal coefficient in a multiresolution wavelet basis using
conditionally independent Dirichlet process priors.
Among the methods that guarantee spatial contiguity, they may
either impose certain constraints on cluster shapes
\citep{knorr2000bayesian,kim2005analyzing,lee2017cluster}, or fail to allow for
globally discontinuous clusters \citep{li2019spatial}.

Finally, an important consideration in clustering is how to determine the
number
of clusters. Most existing methods such as \citet{heaton2017nonstationary}
require specification of the number of clusters first. In Bayesian statistics,
Dirichlet Process mixture models (DPM) have
gained large popularity because of their flexibility in allowing for an unknown
number of clusters. Recently, \cite{miller2018mixture} proved that DPM can
produce an inconsistent estimate of the number of clusters, and proposed a
mixture of finite mixtures model to resolve the issue while inheriting many
attractive mathematical and computational properties of DPM. However, their
method may not be efficient for spatial clustering as it does not take into
account any spatial information.

To address these challenges when facing the analysis of spatial income Lorenz
curves, in this article, we develop a new Bayesian nonparametric method that
combines the ideas of Markov random field models and mixture of finite mixtures
models to leverage geographical information. A distinction of the method is its
ability to capture both locally spatially contiguous clusters and globally
discontiguous clusters. Moreover, it utilizes an efficient Markov chain Monte
Carlo (MCMC) algorithm to estimate the number of clusters and clustering
configuration simultaneously while avoiding complicated reversible jump MCMC or
allocation samplers. We introduce this new Bayesian nonparametric clustering
model to the analysis of the US state level household income Lorenz curves. In
particular, we use a similarity measure among functional curves based on the
inner product matrix under elastic shape analysis
\citep{srivastava2016functional}, which has a nice invariance property to
shape-preserving transformations. The results of real data reveal
interesting clustering patterns of IDs among different states,
which provide important information to study regional income inequalities.


The rest of this paper is organized as follows. The motivating PUMS data is
introduced in detail in Section~\ref{sec:data}. We briefly review 
elastic shape analysis of functions in Section~\ref{sec:function_MFM},
followed by a review of nonparametric Bayesian clustering methods in
Section~\ref{sec:MFM}. We describe the proposed Markov random field constrained
mixture of finite mixture prior model and introduce our functional data
clustering model in Section~\ref{sec:MRFMFM}. In Section~\ref{sec:bayes_comp},
the Bayesian inference including the MCMC sampling algorithm, the  model
selection criterion for tuning parameter, post-MCMC inference, and convergence
diagnostic criteria are introduced. Simulation and case study using the PUMS
data are presented respectively in Sections~\ref{sec:simu}
and~\ref{sec:real_data}.
Section~\ref{sec:discuss} closes the paper with
some conclusions and discussions.

\section{Motivating Data}\label{sec:data}

Our motivating data comes from the 2018 submission in the PUMS data registry.
US
households' incomes and the states they live in are recorded for the 50 states
plus Washington, DC. For simplicity, we refer to them as ``51 states'' in the
rest of this paper. The Lorenz curve \citep{lorenz1905methods} is a commonly
used functional representation of the distribution of income or wealth, which
\blue{reflects} inequality of the wealth distribution. 
It, specifically, assumes that the household
income~$x$ follows a cumulative distribution function (CDF)~$F(x)$ with
respective probability density function~$f(x)$. Let $Q(p)=F^{-1}(p)$ be the
inverse CDF defined as $Q(p) = \inf\{y: F(y)\geq p\}$. The Lorenz curve is
defined as
\begin{equation*}
  L(p) = \frac{1}{\mu}\int_0^p Q(t) \dd t, ~~\mbox{for }0\leq p \leq 1,
\end{equation*}
where $\mu = \int_0^1 Q(t) \dd t$. By definition, when plotted in a graph,
the Lorenz curve always starts
at $(0, 0)$ and ends at (1,1), and measures on the bottom for $100p\%$ of
households, what percentage $100L\%$ of total income they have.

In practice, the empirical Lorenz curve can be constructed from data in a
similar fashion \blue{using the empirical CDF.}
Define
\blue{for state $i$}
\begin{equation*}
  \hat{L}_i(p_k) = \sum_{j=1}^k y_{i, (j)} / \sum_{j=1}^T y_{i, (j)},
\end{equation*}
where $p_k = k/n$, for $k=1,\ldots, n$, and $y_{i,(j)}$ is the $j$-th order
statistic observed in state~$i$. Under mild regularity conditions,
$\hat{L}_i$ converges uniformly in~$p\in[0,1]$ and almost surely to~$L_i$
\citep{gastwirth1972estimation}. The Gini index, as a derived measure, is
defined as two times the area between the Lorenz curve and the 45 degree line
of
equality from~$(0, 0)$ to~$(1,1)$.

\begin{figure}[tbp]
  \center
      \includegraphics[width=0.9\textwidth]{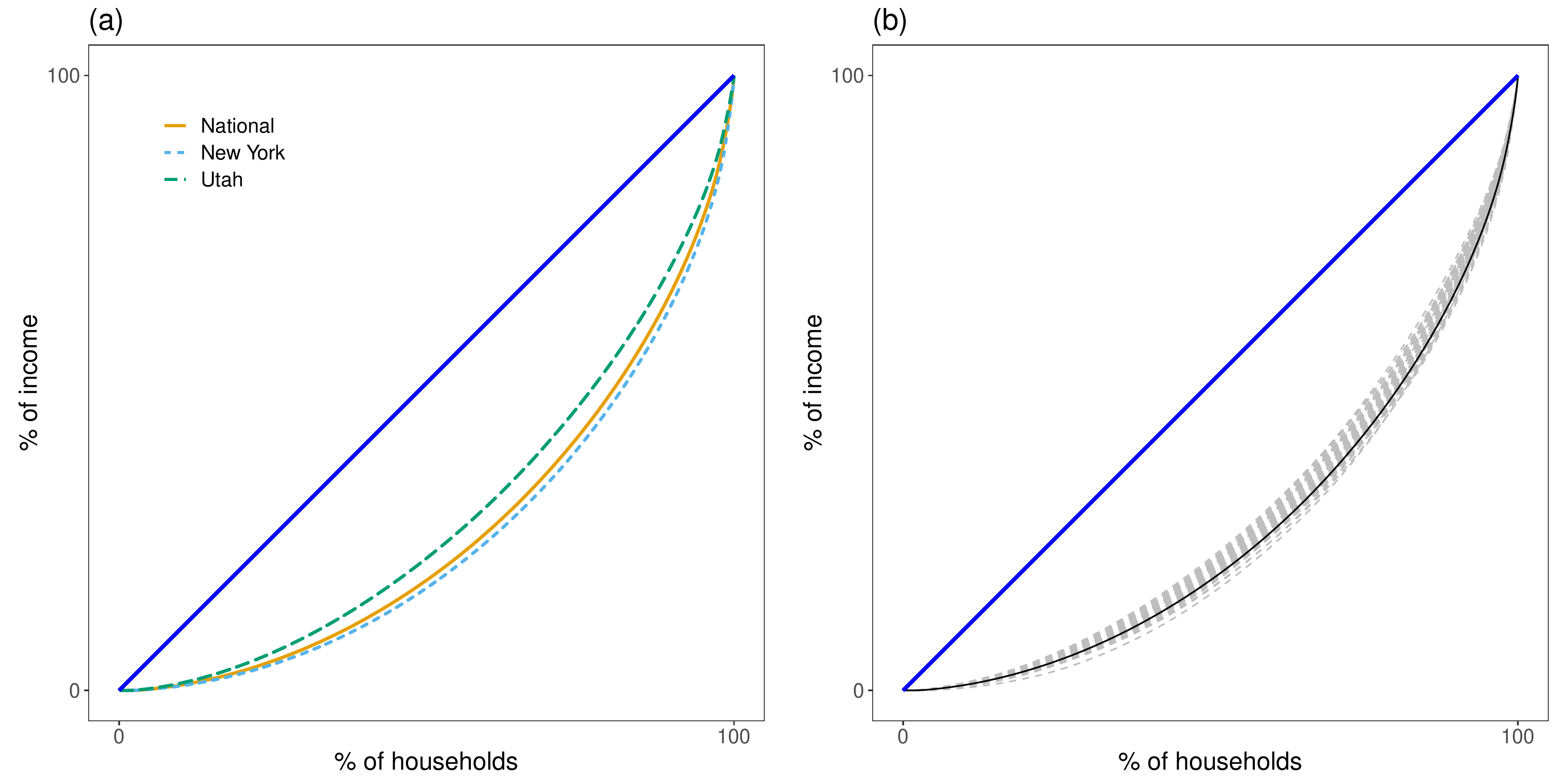}
  \caption{(a) Lorenz curves calculated based on the PUMS 2017 Household Income
  data on the national level and for two selected states; (b) Lorenz curves for
  all US states.}
    \label{fig:lorenz_curve}
  \end{figure}

For the 2017 US household income data, examples of Lorenz curves are presented
in Figure~\ref{fig:lorenz_curve}. The Lorenz curve computed on the national
level using all observations is marked in solid line, with a
corresponding
Gini coefficient of 0.4804. A closer look at the state-level Lorenz curves,
however, reveals that the IDs do vary across states.
The Lorenz curves for two selected states, Utah and New York, are also
illustrated in Figure~\ref{fig:lorenz_curve}(a) as an example. It is rather
apparent that while Utah's curve lies above the national curve, indicating
more equality, New York's curve lies below, suggesting a larger gap between
rich and poor. Lorenz curves for all US states are plotted together with the
national
curve in Figure~\ref{fig:lorenz_curve}(b), and they form a ``cloud''
instead of being similar to each other. The ability of Lorenz curves to
describe income inequalities is clearly demonstrated here.

  \begin{figure}[tbp]
    \centering
    \includegraphics[width = 0.8\textwidth]{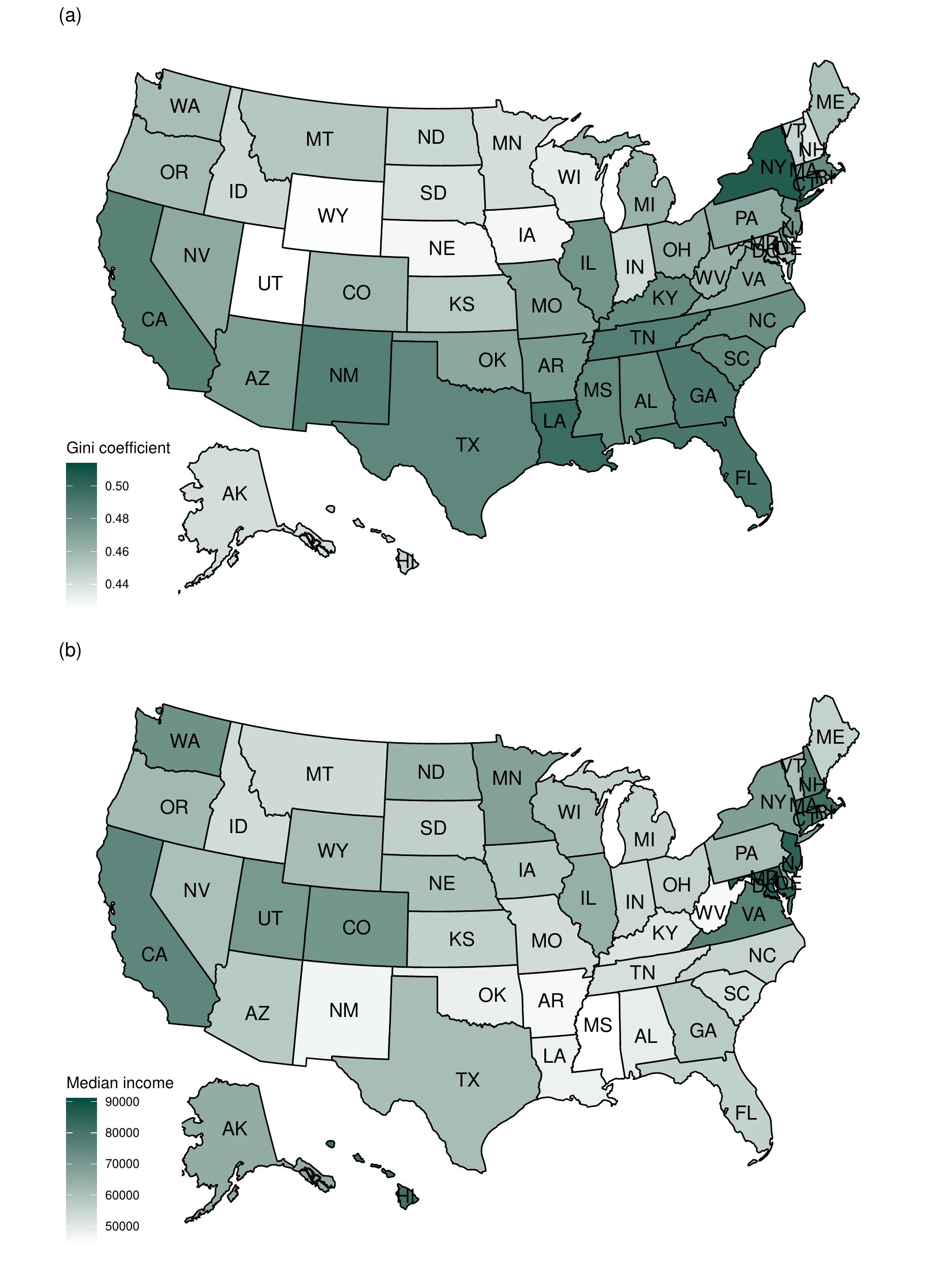}
    \caption{Descriptive statistics of PUMS data on the US map: (a)
Gini coefficient; (b) state median income.}
    \label{fig:realdatadesc}
\end{figure}

In addition to the Lorenz curves, descriptive statistics, which include the
Gini
coefficient and state median income, are presented in
Figure~\ref{fig:realdatadesc}. With a Gini of~0.423, Utah becomes
the state that has the least income inequality, and Washington, DC has the
worst
income inequality with a Gini of 0.512. It also has the highest median income of
\$90,000, while Mississippi has the lowest median income of \$43,500.

\section{Methodology}\label{sec:method}

In this section, we treat the state-level Lorenz curves as spatial functional
data.
We will first discuss the functional representation of ID and the
shape-based similarity measure
\blue{between two IDs}. Next, the nonparametric Bayesian
approach for functional data clustering based on the similarity
measure is introduced.
In addition, a Markov random fields constraint mixture of finite
mixtures model (MRFC-MFM) is proposed to add a spatial constraint in
clustering prior. The hierarchical model under MRFC-MFM is presented at the end
of this section.

\subsection{Functional Representation of Income
Distribution}\label{sec:function_MFM}

We begin the section by reviewing the functional data shape analysis technique.
In order to cluster functional data, we need to define appropriate metrics to
quantify similarities among functional curves. There are four important
features
of functional data including quantity, frequency, similarity, and smoothness.
Commonly used distance metrics such as the Euclidean distance are no longer
appropriate candidates for quantifying similarities between functions. In this
article, we
consider the inner product matrix calculated using a specific representation of
curves called the square-root velocity function
\citep[SRVF;][]{srivastava2010shape}. This inner product matrix is a summary
statistic that encodes the similarity information among curves for subsequent
clustering analysis. This inner product matrix will focus more on the
differences between the shape of functions. By focusing on shapes, one is more
interested in the numbers and relative heights of peaks and valleys in a curve,
rather than their precise locations. This property will be more suitable for
quantifying the differences of IDs among different regions, because the precise
locations or mean shifts have less effects on inequality of ID.

The SRVF of an absolutely continuous function $f(t): [0,1] \rightarrow
\mathcal{R}^{p}$ is defined as:
\begin{equation}
q(t)=\text{sign}\big(f'(t)\big)\sqrt{|f'(t)|},  
\label{eq:srvf_def}
\end{equation}
where $f'(t)$ is the first order derivative of function $f$ on $t$.
\blue{It can be seen that the SRVF is a curve of unit length.} There are
several advantages of using SRVF for functional data analysis. First, the
scaling, rotation and re-parameterization variabilities still remain based on
SRVF. In addition, the elastic metric is invariant to the reparameterization of
functions. The SRVF represents unit-length curves as a unit hypersphere in the
Hilbert manifold. The SRVF for a given function can be obtained in \textsf{R}
using the \texttt{f\_to\_srvf()} function provided by the \textbf{fdasrvf}
package \citep{Rpkg:fdasrvf}. For given functions~$f_1$ and~$f_2$ which belong
to~$\mathcal{F}=\{f: [0,1]\rightarrow \mathcal{R}^p: \text{$f$ is absolutely
continuous}\}$ and their corresponding SRVFs, $q_1$ and $q_2$, the inner
product is defined  based on the definition in \cite{zhang2015bayesian} as
follows:
\begin{equation}
S_{f_1,f_2}=\sup_{\gamma\in \Gamma,O \in SO(p)}\big\langle
q_1,\big(q_2,(O,\gamma)\big)\big\rangle,
\label{eq:inner_product}
\end{equation}
where $SO(p)$ \blue{is the collection of
orthogonal~$p\times p$, i.e. $p$-dimensional
rotation matrices,
and~$\Gamma$ represents the set of all orientation-preserving
diffeomorphisms over the domain~[0,1]. The notation~$(O,\gamma)$
denotes a joint action of the rotation and reparameterization
operations, and~$\big(q2,(O,\gamma)\big)$ here represents certain
reparameterizations and rotations of~$q_2$.  The maximization
over~$SO(p)$ and~$\Gamma$ can be performed iteratively as in
\citet{srivastava2010shape}.
The operation $\langle \cdot, \cdot \rangle$ denotes the inner product in
$\mathbb{L}^2([0,1], \mathbb{R}^p)$: $\langle v, q \rangle = \int_0^1 \langle
v(t), q(t)\rangle \dd t$. The value of the integral ranges from~$-1$ to~1,
with~$-1$ indicating that the curves are exactly the opposite, and~1 indicating
that they are exactly the same.}
The inner product of two functions can be calculated using the algorithm in
\cite{tucker2013generative}. Computation in~\textsf{R} is facilitated with the
\texttt{trapz()} function in package
\textbf{pracma} \citep{Rpkg:pracma}. Given $f_1,\ldots, f_n$, the~$n\times n$
pairwise inner product matrix $\bm{S}$ can be calculated using the definition
in~\eqref{eq:inner_product}, and \textbf{fdasrvf} and \textbf{pracma}.


\blue{\subsection{Mixture of Finite Mixtures for Distances Between Functional
Data}}\label{sec:MFM}
Next, we introduce nonparametric Bayesian methods to capture spatial
homogeneity
of functional data. We start with a Fisher's $Z$-transformation of the inner
product matrix~$\bm{S}$ to make each entry~$\bm{S}_{f_i,f_j}$ of the matrix
within the range of a Gaussian distribution. The transformed inner product
matrix is denoted as~$\bm{\mathscr{\bm{S}}}$, with each entry being
\begin{equation*}
  \mathscr{\bm{S}}_{ij} = \log\left(
  \dfrac{1 + \bm{S}_{f_i, f_j}}{1 - \bm{S}_{f_i, f_j}}
  \right).
\end{equation*}
%
The larger~$\mathscr{S}_{ij}$ is, the closer $f_i$ and $f_j$ are. We further
assume that
\begin{equation}\label{eq:sBm}
\begin{split}
\mathscr{S}_{ij} &\mid \bm{\mu}, \bm{\tau}, k \sim
\mbox{N}(\mu_{ij},\tau_{ij}^{-1}),
\quad \mu_{ij} = U_{z_i z_j}\\
\tau_{ij} &= T_{z_i z_j}, \quad 1 \leq i \leq j
\leq n,
\end{split} 
\end{equation}
where $k$ is the number of true underlying clusters, $\mbox{N}()$ denotes the
normal distribution, $z_i \in \{1, \ldots, k\}$ denotes the cluster membership
of the $i$-th curve; $\bm{U} = [U_{rs}] \in (-\infty,+\infty)^{k \times k}$ and
$\bm{T} = [T_{rs}] \in (0,+\infty)^{k \times k}$ are symmetric matrices, with
$U_{rs} = U_{sr}$ indicating the mean closeness of any function $f_i$ in
cluster
$r$ and any function $f_j$ in cluster $s$, and~$T_{rs} = T_{sr}$ indicating the
precision of closeness between any function $f_i$ in cluster $r$ and any
function $f_j$ in cluster $s$. Note that in the above formulation, only the
upper triangle of matrix $\bm{\mathscr{S}}$ is modeled, including the diagonal.

Let $\m Z_{n, k} = \big\{(z_1, \ldots, z_n) : z_i \in \{1, \ldots, k\}, 1 \le i
\le n \big\}$ denote all possible partitions of~$n$ nodes into~$k$ clusters.
Given~$z \in \m Z_{n, k}$, let~$\bm{\mathscr{\bm{S}}_{[rs]}}$ denote the~$n_r
\times n_s$ sub-matrix of~$\bm{\mathscr{S}}$ consisting of entries
$\mathscr{\bm{S}}_{ij}$ with~$z_i = r$ and~$z_j = s$.
\blue{Following the common practice for
stochastic block models \citep[SBM;][]{holland1983stochastic},
independence between entries of~$\bm{\mathscr{S}}$, or edges, is assumed.}
The joint likelihood of
$\bm{\mathscr{S}}$ under model~\eqref{eq:sBm} can be expressed as 
\begin{equation}\label{eq:like}
\begin{split}
P(\bm{\mathscr{S}} \mid \bm{z}, \bm{U}, \bm{T}, k) &= \prod _{1 \leq r\leq s
\leq k}
P(\bm{\mathscr{S}_{[rs]}}\mid
\bm{z}, \bm{U},\bm{T}), \\
P(\bm{\mathscr{S}_{[rs]}} \mid \bm{z}, \bm{U},\bm{T}) &= \prod_{1\leq i < j
\leq n:
z_i = r, z_j = s} \frac{1}{\sqrt{2\pi T_{rs}^{-1}}}
\exp\left\{-\frac{T_{rs}(\mathscr{S}_{ij}-U_{rs})^2}{2}\right\}.
\end{split}	
\end{equation}
A common Bayesian specification when $k$ is given can be completed by assigning
independent priors to $\bm{z}$, $\bm{U}$ and $\bm{T}$, and it can be easily
incorporated into a framework as finite mixture models. A popular solution for
unknown $k$ is to introduce the Dirichlet process mixture prior models
\citep{antoniak1974mixtures} as following: 
\begin{eqnarray}\label{eq:DPMM}
\bm{\mathscr{S}}_i \sim   F(\cdot ,\bm{\theta}_i),
\quad \bm{\theta}_i \sim  G(\cdot),
\quad G \sim  DP(\alpha G_0),
\end{eqnarray}
where $\bm{\mathscr{S}_i} =
(\mathscr{S}_{i1},\mathscr{S}_{i2},\ldots,\mathscr{S}_{in})$,
$\bm{\theta}_i = (\bm{\theta}_{i1},\bm{\theta}_{i2},\ldots,\bm{\theta}_{in})$
and $\bm{\theta}_{ij} = (\mu_{ij},\tau_{ij})$.

Dirichlet process is parameterized by a base measure $G_0$ and a concentration
parameter~$\alpha$. If a set of values of $\bm{\theta}_1,\ldots,\bm{\theta}_n$
are drawn
from $G$, a conditional prior can be obtained by integration
\citep{blackwell1973ferguson}:
\begin{eqnarray}\label{eq:DPMM1}
p(\bm{\theta}_{n+1}\mid \bm{\theta}_1,\ldots,\bm{\theta}_n) =
\dfrac{1}{n+\alpha}\sum_{i=1}^n\delta_{\bm{\theta}_i}(\bm{\theta}_{n+1}) +
\dfrac{\alpha}{n+\alpha}G_0(\bm{\theta}_{n+1}).
\end{eqnarray}
Here, $\delta_{\bm{\theta}_i}(\bm{\theta}_{j}) = I(\bm{\theta}_j =
\bm{\theta}_i)$ is the distribution concentrated at a single point
$\bm{\theta}_i$. Equivalent models can also be obtained by introducing cluster
membership $z_i$'s and letting the
unknown number of clusters $K$ go to infinity \citep{neal2000markov}. 
\begin{equation}\label{eq:DPMM2}
\begin{split}
\bm{\mathscr{S}}_i \mid z_i, \bm{\theta}^* & \sim F(\bm{\theta}^*_{z_i}),\\
z_i \mid  \bm{\pi} & \sim \text{Discrete} (\pi_1,\ldots,\pi_K), \\
\bm{\theta}^*_c & \sim G_0\\
\bm{\pi} & \sim \text{Dirichlet}(\alpha/K,\ldots ,\alpha/K) 
\end{split}
\end{equation}
where $\bm{\pi}=(\pi_1,\ldots,\pi_K)$. For each cluster $c$, the parameters
$\bm{\theta}^*_c$ determine the cluster specific distribution~$F(\cdot \mid
\bm{\theta}^*_c)$.

By integrating out mixing proportions $\bm{\pi}$, we can obtain the prior
distribution of $(z_1, z_2, \ldots, z_n)$ that allows for automatic inference
on
the number of clusters $k$, which is also well known as the Chinese restaurant
process
\citep
[CRP;][]{aldous1985exchangeability,pitman1995exchangeable,neal2000markov}.
Through the popular Chinese restaurant metaphor, $z_i$, $i=2, \ldots, n$ are
defined through the following conditional distribution \citep[P\'{o}lya urn
scheme,][]{blackwell1973ferguson}:
\begin{eqnarray}\label{eq:crp}
P(z_{i} = c \mid z_{1}, \ldots, z_{i-1})  \propto   
\begin{cases}
\abs{c}  , &  \text{at an existing table labeled}\, c\\
\alpha,  & \text{if} \, $c$\,\text{is a new table}
\end{cases},
\end{eqnarray}
where $\abs{c}$ is the size of cluster $c$.

While the CRP has a very attractive feature of simultaneous estimation on the
number of clusters and the cluster configuration, a striking limitation of this
model has been recently discovered. \cite{miller2018mixture} proved that the
CRP
produces extraneous clusters in the posterior leading to inconsistent
estimation
of the {\em number of clusters} even when the sample size grows to infinity. A
modification of the CRP called mixture of finite mixtures (MFM) model is
proposed to circumvent this issue \citep{miller2018mixture}:
\begin{eqnarray}\label{eq:MFM}
k \sim p(\cdot), \quad (\pi_1, \ldots, \pi_k) \mid k \sim \mbox{Dirichlet}
(\gamma,
\ldots, \gamma), \quad z_i \mid k, \bm{\pi} \sim \sum_{h=1}^k  \pi_h
\delta_h,\quad
i=1, \ldots, n, 
\end{eqnarray}
where~$p(\cdot)$ is a proper probability mass function (p.m.f.) on~$\{1, 2,
\ldots, \}$ and~$\delta_h$ is a point-mass at~$h$. Compared to the CRP, the
introduction of new tables is slowed down by the factor~$V_n(w+1)/ V_n(w)$,
which facilitates a model-based pruning of the tiny extraneous clusters. The
coefficient~$V_n(w)$ needs to be precomputed as:
\begin{equation*}
V_n(w) = \sum_{k=1}^{+\infty}\dfrac{k_{(w)}}{(\gamma k)^{(n)}} p(k),
\end{equation*}
where $k_{(w)}=k(k-1)\ldots(k-w+1)$, and $(\gamma k)^{(n)} = {\gamma k}(\gamma
k+1)\ldots(\gamma k+n-1)$. By convention,~$x^{(0)} = 1$ and $x_{(0)}=1$.

The conditional prior of $\bm{\theta}$ under MFM can be stated as below:
\begin{eqnarray}\label{eq:MFM1}
P(\bm{\theta}_{n+1}\mid \bm{\theta}_1,\ldots,\bm{\theta}_n) \propto
\sum_{i=1}^w(n_i + \gamma)\delta_{\bm{\theta}^*_i} +
\dfrac{V_n(w+1)}{V_n(w)}\gamma G_0(\bm{\theta}_{n+1}).
\end{eqnarray}
where $\bm{\theta}^*_1,\dots,\bm{\theta}^*_w$ are the distinct values taken by
$\bm{\theta}_1,\dots,\bm{\theta}_n$ and $w$ is the number of existing clusters.
The cluster membership $z_i$, for $i=2, \ldots, n$, in \eqref{eq:MFM} can be
defined in a P\'{o}lya urn scheme similar to CRP:
\begin{eqnarray}\label{eq:mcrp}
P(z_{i} = c \mid z_{1}, \ldots, z_{i-1})  \propto   
\begin{cases}
\abs{c} + \gamma  , &  \text{at an existing table labeled }\, c\\
V_n(w+1)/ V_n(w)\gamma,  & \text{if } \, c\,\text{ is a new
table}
\end{cases}.
\end{eqnarray}
where $w$ is the number of existing clusters.   

   
Adapting MFM to our model setting for functional clustering, the model and
prior
can be expressed hierarchically as: 
\begin{align}\label{eq:Func_MFM}
& k \sim p(\cdot), \quad \text{where $p(\cdot)$ is a p.m.f on \{1,2, \ldots\} }
\nonumber \\
& T_{rs} = T_{sr} \stackrel{\text{ind}} \sim \mbox{Gamma}(\alpha, \beta),  
\quad r, s = 1, \ldots, k,  \nonumber
\\ \nonumber
& U_{rs} = U_{sr} \stackrel{\text{ind}} \sim \mbox{N}(\mu_0, k_0^{-1} 
T_{rs}^{-1}),  \quad r, s = 1, \ldots, k,  \nonumber
\\
& \mbox{pr}(z_i = j \mid \bm{\pi}, k) = \pi_j, \quad j = 1, \ldots, k, \, \quad
i =
1,
\ldots, n,  \\ \nonumber
& \bm{\pi} \mid k \sim \mbox{Dirichlet}(\gamma, \ldots, \gamma),\\
& \mathscr{\bm{S}}_{ij} \mid \bm{z}, \bm{U},\bm{T}, k \stackrel{\text{ind}}
\sim
\mbox{N}(\mu_{ij}, 
\tau_{ij}^{-1}), \quad \mu_{ij} = U_{z_i z_j},~~ \tau_{ij} = T_{z_i z_j},~~ 1
\leq i < j \leq n.  \nonumber
\end{align}
We assume $p(\cdot)$ is a $\mbox{Poisson}(1)$ distribution truncated to be
positive through the rest of the paper, which has been proved by
\cite{miller2018mixture} and \cite{geng2019probabilistic} to guarantee
consistency for the mixing distribution and the number of clusters.  We refer
to
the hierarchical model in \eqref{eq:Func_MFM} as MFM-fCluster.

\subsection{Markov Random Field Constrained MFM in Functional
data}\label{sec:MRFMFM}

A possible weakness of MFM for spatial functional data is due to its inability
to account for spatial structure or dependence, i.e., MFM neglects the
spatial smoothness of a map, and hence the resulting clustering does not comply
any spatial constraints, and therefore
might be sensitive to noise in the data. This
drawback can be addressed by introducing spatial coupling between adjacent
features.  Applying a Markov random field prior to spatial statistical modeling
is a classical Bayesian approach widely used in image segmentation problems
\citep{geman1984stochastic}. In this section, we combine the similar idea of
Markov random fields with MFM to introduce spatial constraints for clustering.

The Markov random field \citep[MRF;][]{orbanz2008nonparametric} provides a
convenient approach to address the difficult problem of modeling a collection
of
dependent random variables \citep{winkler2012image}. \blue{The dependence
structure of different states can be represented by a graph, with vertices
representing random variables and an edge between two vertices indicating
statistical dependence, which can conveniently introduce the spatial
smoothness for both Gaussian and non-Gaussian data (e.g., mixture data).}
Interactions among
variables are constrained to a small group that are usually assumed to be
closer
spatially, in order to reduce the complexity of the problem. The neighborhood
dependence structure of MRF is encoded by a weighted graph $\mathscr{N} =
(V_{\mathscr{N}},E_{\mathscr{N}},W_{\mathscr{N}})$ in space, with vertices
$V_\mathscr{N} = (v_1,\ldots,v_n)$ representing random variables at~$n$ spatial
locations,  $E_\mathscr{N}$ denoting a set of edges representing statistical
dependence among vertices, and $W_{\mathscr{N}}$ denoting the edge weights
representing the magnitudes of dependence.

The MRF for a collection of random variables
$\bm{\theta}_1,\ldots,\bm{\theta}_n$ on a graph $\mathscr{N}$ has a valid joint
distribution $M(\bm{\theta}_1,\ldots,\bm{\theta}_n) :=
\frac{1}{Z_H}\text{exp}\{-H(\bm{\theta}_1,\ldots,\bm{\theta}_n)\}$, with~$H$
being the cost function with the following form
\begin{eqnarray}\label{eq:cost}
H(\bm{\theta}_1,\ldots,\bm{\theta}_n) := \sum_{A \in \mathscr{C}_N}
H_A(\bm{\theta}_A),
\end{eqnarray}
where $\mathscr{C}_N$ denotes the set of all cliques in $\mathscr{N}$, each
term
$H_{A}$  is a non-negative function over the variables in clique $A$, and $Z_H$
is a normalization term. By Hammersley-Clifford theorem, the corresponding
conditional distributions enjoy the Markov property, i.e., $M(\bm{\theta}_i\mid
\bm{\theta}_{-i})=M(\bm{\theta}_i \mid \bm{\theta}_{\partial(i)})$,
where~$\partial(i) := \{j\mid (i,j) \in E_\mathscr{N}\}$ denotes the set of
neighbors of observation~$i$. Considering only pairwise interactions, we model
the conditional cost functions as 
\begin{eqnarray}\label{eq:costuse}
H(\bm{\theta}_i\mid\bm{\theta}_{-i}) := -\lambda \sum_{l\in \partial(i)}
I(\bm{\theta}_l =
\bm{\theta}_i) = -\lambda\sum_{l\in \partial(i)} I(z_l = z_i),
\end{eqnarray}
where \blue{$\lambda \in \mathbb{R}^+$} is a parameter controlling the
magnitude of spatial smoothness, \blue{with larger values inducing stronger
spatial smoothing. It can be seen that the function takes value
in~$\{0,-\lambda\}$.}


Markov random field constrained MFM (MRFC-MFM) is composed of an interaction
term modeled by a MRF cost function to capture spatial interactions among
vertices and a vertex-wise term modeled by a MFM. The resulting model defines a
valid MRF distribution~$\Pi$, which can be written as
\begin{equation}\label{eq:MRFMFM1}
\Pi(\bm{\theta}_1,\ldots,\bm{\theta}_{n}) \propto
P(\bm{\theta}_1,\ldots,\bm{\theta}_{n})M(\bm{\theta}_1,\ldots,\bm{\theta}_{n})
\end{equation}
with $P(\bm{\theta}_1,\ldots,\bm{\theta}_{n})$ defined by the conditional
distributions in \eqref{eq:MFM1} and $M(\bm{\theta}_1,\ldots,\bm{\theta}_{n})$
from the MRF model using \eqref{eq:costuse} as the conditional cost function.
This constrained model exhibits a key
property that the MRF constraints only change the finite component of the MFM
model as shown in Theorem \ref{thm:MRF-MFM} below. The proof is deferred to
Appendix \ref{sec:proof}. 

\begin{thm}\label{thm:MRF-MFM} Let $n_k^{(-i)}$ denote the size of the $k$-th
cluster excluding $\theta_i$, $K^*$ denote the number of clusters excluding the
$i$-th observation, and assume $H(\theta_i\mid\theta_{-i})$ is a valid MRF
conditional cost
function. The  conditional distribution of a MRFC-MFM takes the form
\begin{equation*}
\Pi(\bm{\theta}_i\mid\bm{\theta}_{-i}) \propto \sum_{k=1}^{K^*}
(n_k^{(-i)}+\gamma)\frac{1}{Z_H}\exp(-H(\bm{\theta}_i\mid
\bm{\theta}_{-i}))\delta_{\bm{\theta}^*_k}
(\bm{\theta}_{i}) + \dfrac{V_n(K^*+1)}{V_n(K^*)}\dfrac{\gamma}
{Z_H}G_0(\bm{\theta}_{i}).
\end{equation*}
\end{thm}
An immediate corollary of Theorem~\ref{thm:MRF-MFM} can be defined in a
P\'{o}lya urn scheme after introducing the cluster assignments parameters
$z_i$,
$i=1, \ldots, n$. 

\begin{corollary}\label{cor:MRF-MFM}
Suppose the conclusion of Theorem \ref{thm:MRF-MFM} holds. Then, 
\begin{align*}
\Pi(z_{i} = c \mid \bm{z}_{-i})  \propto   
\begin{cases}
\left[\abs{c} + \gamma\right] \exp[\lambda \sum_{l\in \partial(i)} I(z_l =
z_i)]  , &
\text{at
an existing table labeled }\, c\\
V_n(K^*+1)/ V_n(K^*)\gamma,  &  \text{if } \, c\,\text{ is a new
table}
\end{cases},
\end{align*}
where~$\bm{z}_{-i} = \bm{z} \backslash \{z_i\}$, i.e., all elements of
$\bm{z}$
except for~$z_i$. 
\end{corollary}

The above scheme offers an intuitive interpretation of MRFC-MFM again using the
Chinese restaurant metaphor: the probability of a customer $i$ sitting at a
table depends not only on the number of other customers already sitting at that
table, but also the number of other customers that have spatial ties to the
$i$-th customer. The parameter~$\lambda$ controls the strength of spatial ties,
and ultimately controls estimation on the number of clusters. The larger the
value for~$\lambda$, the stronger the spatial smoothing effect and the smaller
the number of clusters. This can be clearly observed in the simulation results
presented in the sensitivity analysis section of the supplemental material. In
particular, the MFM model developed in \cite{miller2018mixture}  can be viewed
as a special case of MRFC-MFM when~$\lambda=0$. We use the notation
$\text{MRFC-MFM}(\lambda,G_0)$ to represent the MRFC-MFM prior with a
smoothness
parameter $\lambda$ and a base distribution $G_0$.  The Markov random field
constraint-mixture of finite mixture-functional clustering method
(MRFC-MFM-fCluster) can be hierarchically written as 
\begin{align}\label{eq:Func_MRFMFM}
&\bm{U},\bm{T},\bm{z},k \sim \text{MRFC-MFM}(\lambda,G_0),\\
& \mathscr{\bm{S}}_{ij} \mid \bm{z}, \bm{U},\bm{T}, k \stackrel{\text{ind}}
\sim
\mbox{N}(\mu_{ij}, 
\tau_{ij}^{-1}), \quad \mu_{ij} = U_{z_i z_j}, \tau_{ij} = T_{z_i z_j} \quad 1
\leq i < j \leq n,  \nonumber
\end{align}
where~$G_0$ is a normal-gamma distribution whose hyperparameters are the same
with \eqref{eq:Func_MFM}. It is noted that while the model in
\eqref{eq:Func_MRFMFM} introduces spatial dependence to promote locally
contiguous clustering, it still allows any customer a chance to sit with any
other customer so that globally discontiguous clustering can be captured.

\section{Bayesian Inference}\label{sec:bayes_comp}

MCMC is used to draw samples from the posterior distributions of the model
parameters. In this section we present the sampling scheme, the posterior
inference of cluster configurations, and metrics to evaluate the estimation
performance and clustering accuracy.

\subsection{The MCMC Sampling Schemes}\label{sec:mcmc_sampling}

Our goal is to sample from the posterior distribution of the unknown parameters
$k$, $\bm{z} = (z_1, \ldots, z_n) \in \{1, \ldots, k\}^n$, $\bm{U} = [U_{rs}]
\in (-\infty,+\infty)^{k \times k}$ and $\bm{T} = [T_{rs}] \in (0,+\infty)^{k
\times k}$. While inference in MFMs can be done with methods such as reversible
jump Markov chain Monte Carlo or even allocation samplers, they often suffer
from poor mixing and slow convergence. We adapt the algorithm in
\cite{miller2018mixture} to exploit the P\'{o}lya urn scheme for the MRFC-MFM.
An efficient collapsed Gibbs sampler is used for Bayesian inference by
marginalizing out $k$ analytically. The sampler for MFM is presented in
Algorithm~1 in the supplemental material, and the sampler for MRFC-MFM is
presented in Algorithm~2 in the supplemental material. These two algorithms
only
differ by the posterior probability of an observation assigned to an existing
cluster. Both algorithms efficiently cycle through the full conditional
distributions of $z_i$ given $\bm{z}_{-i}$, $\bm{U}$, and~$\bm{T}$ for~$i=1, 2,
\ldots, n$.

For the hyperparameters in both simulation studies and real data analysis, we
use $\alpha=1$, $\beta=1$, $k_0=2$ and $\gamma = 1$.  For $\mu_0$, $\max_{i,j}
\mathcal{S}$ is assigned to diagonal terms and $\min_{i,j} \mathcal{S}$ is
assigned to off-diagonal terms in order to make it more informative. The
choices
of~$\mu_0$ ensure that the functions within a cluster are closer to each
other than between clusters. We arbitrarily initialized the algorithms with
nine
clusters, and randomly allocated the cluster configurations. Various other
choices were tested and we did not find any evidence of sensitivity to the
initialization.

\subsection{Post MCMC Inference}\label{sec:sum_mcmc}

Dahl's method \citep{dahl2016} is a popular post-MCMC inference algorithm for
the clustering configurations $\bm{z}$ and the estimated parameters. The
inference of Dahl's method is based on the membership matrices,
$B^{(1)},\ldots,B^{(M)}$, from the posterior samples. The  membership matrix
$B^{(t)}$ for the $t$-th post-burn-in MCMC iteration is defined as:
\begin{align}
B^{(t)} = [B^{(t)}(i,j)]_{i,j\in \{1:n\}} = 1(z_i^{(t)} = z_j^{(t)})_{n\times
n},~~t=1,\ldots,M,
\end{align}
where $1()$ denotes the indicator function, i.e., $B^{(t)}(i,j)=1$ indicates
observations~$i$ and~$j$ are in the same cluster in the $t$-th posterior sample
after burn-in iterations. Based on the membership matrices for the posterior
samples, an Euclidean mean for membership matrices is calculated by:
\begin{equation*}
  \bar{B} =\frac{1}{M} \sum_{t=1}^M B^{(t)}.
\end{equation*}
The iteration with the least squared distance to $\bar{B}$ is obtained by
\begin{align}
C_{LS} = \text{argmin}_{t \in (1:M)} \sum_{i=1}^n \sum_{j=1}^n \{B(i,j)^{(t)} -
\bar{B}(i,j)\}^2.
\end{align}  
The estimated parameters, together with the cluster assignments~$\bm{z}$, are
then extracted from the~$C_{LS}$-th post burn-in iteration. An advantage of the
Dahl's method is to utilize the information of the empirical pairwise
probability matrix~$\bar{B}$.

Convergence diagnostic of the clustering algorithm is evaluated using the
Adjusted Rand index \citep[ARI;][]{hubert1985comparing}. As an adjusted version
of the Rand Index \citep[RI;][]{rand1971objective}, it measures the concordance
between two clustering schemes, after accounting for chances. Taking values
between 0 and 1, a large ARI value indicates high concordance. In particular,
when two cluster configurations are identical in terms of modulo labeling of
nodes, the ARI takes value~1.

\subsection{Selection of $\lambda$}
In our MRFC-MFM-fCluster algorithm, it is rather important to choose an
appropriate value for~$\lambda$, which controls the magnitude of spatial
smoothness. Under the Bayesian framework, the deviance information criterion
\citep[DIC;][]{spiegelhalter2002bayesian}, the Bayesian equivalent of the
Akaike
information criterion \citep[AIC;][]{akaike1973information}, has been one of
the
most frequently used model selection criterion. The AIC, however, does not
exert
enough penalization for clustering problems, which often leads to
over-clustering
results. Therefore, we consider using a modified version of DIC (mDIC), which
modifies the magnitude of the penalty term of classic DIC to be the same as the
Bayesian information criterion \citep[BIC;][]{schwarz1978estimating}. The mDIC
is calculated as
\begin{equation}
    \text{mDIC}=\text{Dev}({\bar{\theta}})+\log(\frac{n\times(n+1)}{2})p_{D},
\end{equation}
where $$\text{Dev}({{\theta}})=-2\log \prod_{1\leq i < j
\leq n:
z_i = r, z_j = s} \frac{1}{\sqrt{2\pi T_{rs}^{-1}}}
\exp\left\{-\frac{T_{rs}(\mathscr{S}_{ij}-U_{rs})^2}{2}\right\},$$
$\theta=\{1\leq i < j \leq n:
z_i = r, z_j = s, U_{rs},T_{rs}\}$, and
$$p_{D}={\overline {D(\theta )}}-D({\bar {\theta }}),$$
with
$\bar{\theta}$ being the estimated parameters based
on Dahl's method. The model with smaller value of mDIC is preferred.

\section{Simulation}\label{sec:simu}

In this section, we detail the simulation settings, the evaluation metrics, and
the comparison performance results. 

\subsection{Simulation Setting and Evaluation Metrics}\label{ssec:setting}

\begin{figure}[tbp]
  \centering
  \includegraphics[width = 0.5\textwidth]{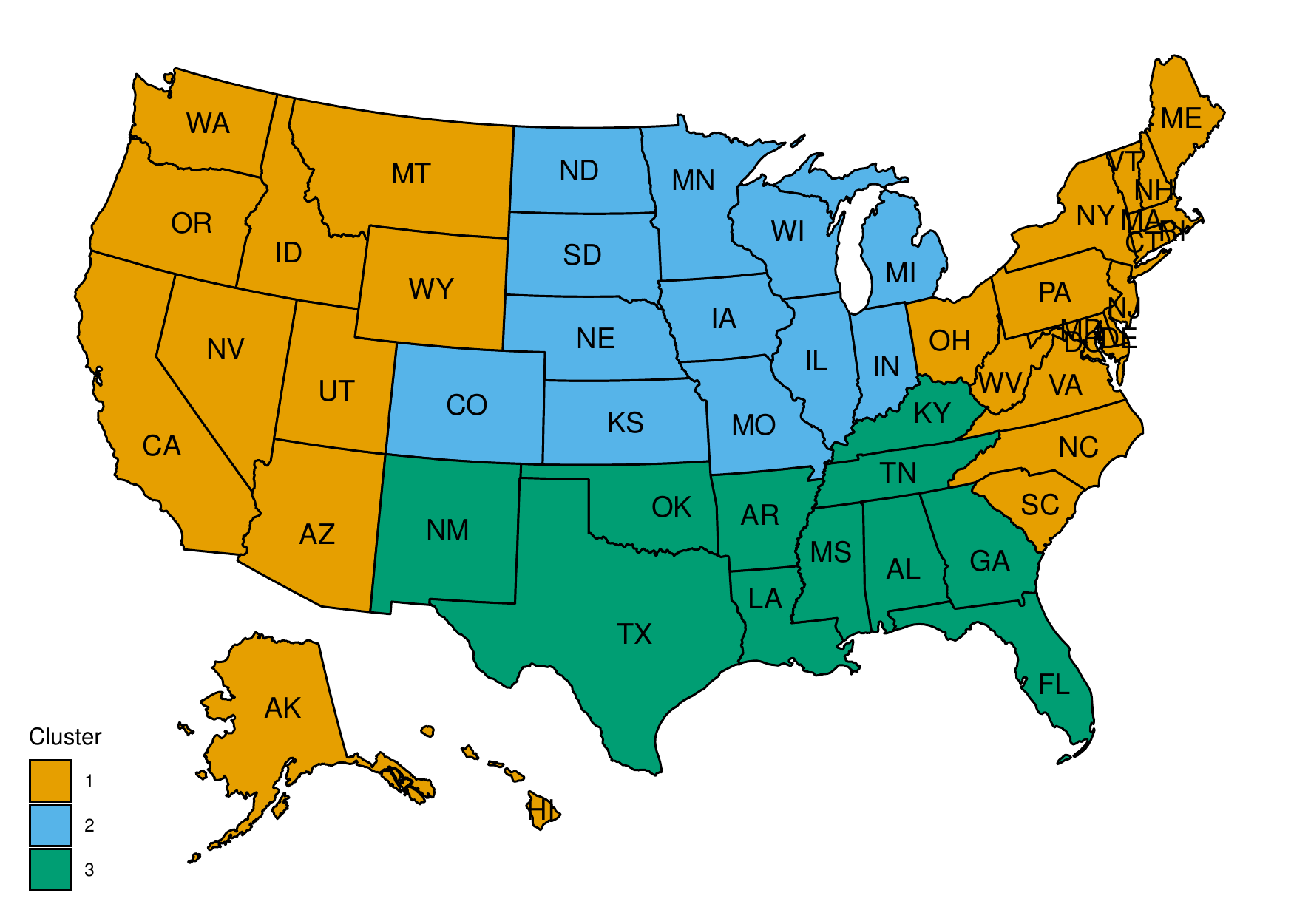}
  \caption{Illustration of the first partition setting with three true
clusters,
  where the first cluster consists of two disjoint components.}
  \label{fig:3clusterdesign}
\end{figure}

\begin{figure}[tbp]
  \centering
  \includegraphics[width = 0.5\textwidth]{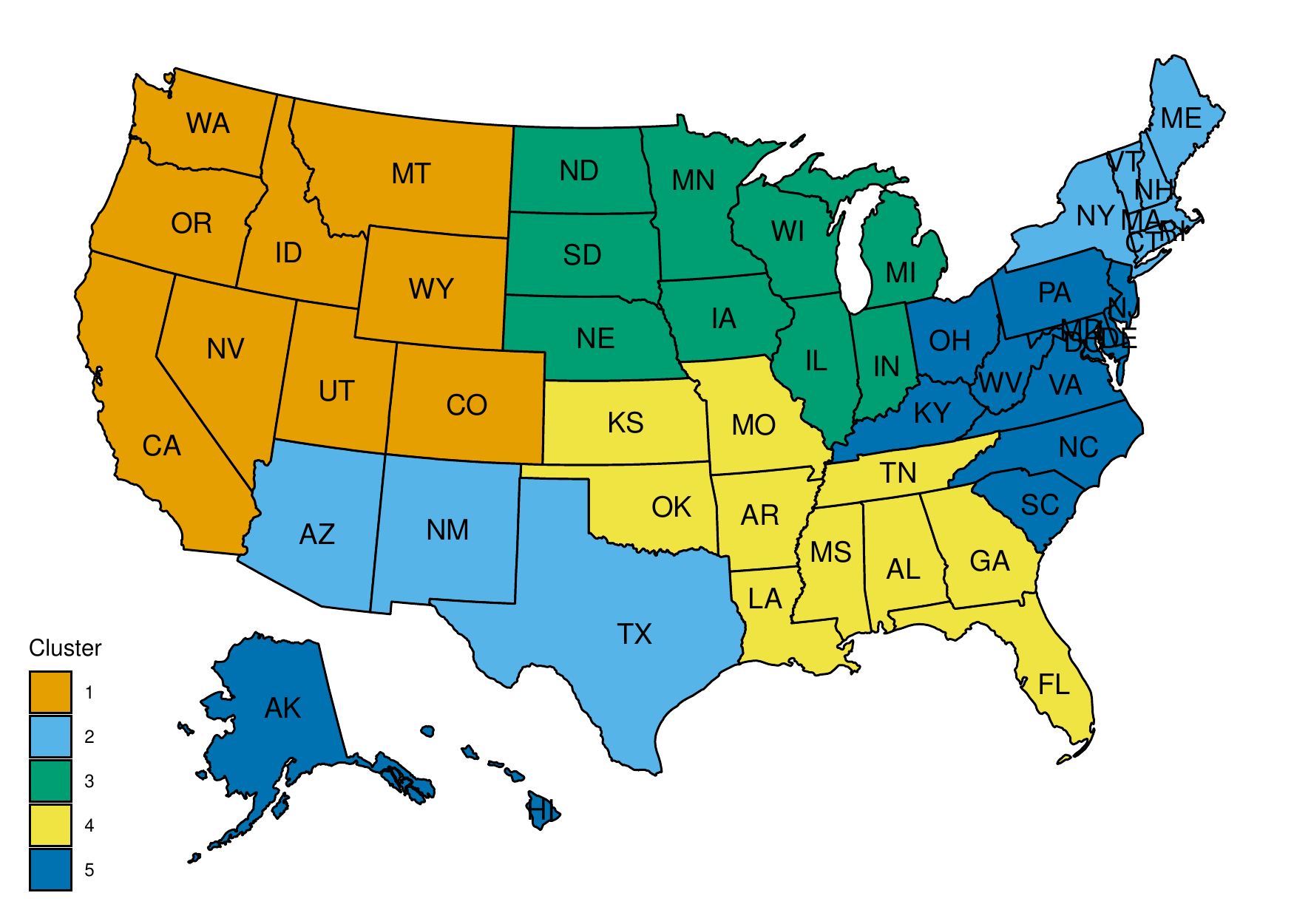}
  \caption{Illustration of the second partition setting with five clusters,
  where clusters~2 and 5 both have disjoint components.}
  \label{fig:5clusterdesign}
\end{figure}

\begin{figure}[tbp]
  \centering
  \includegraphics[width = 0.5\textwidth]{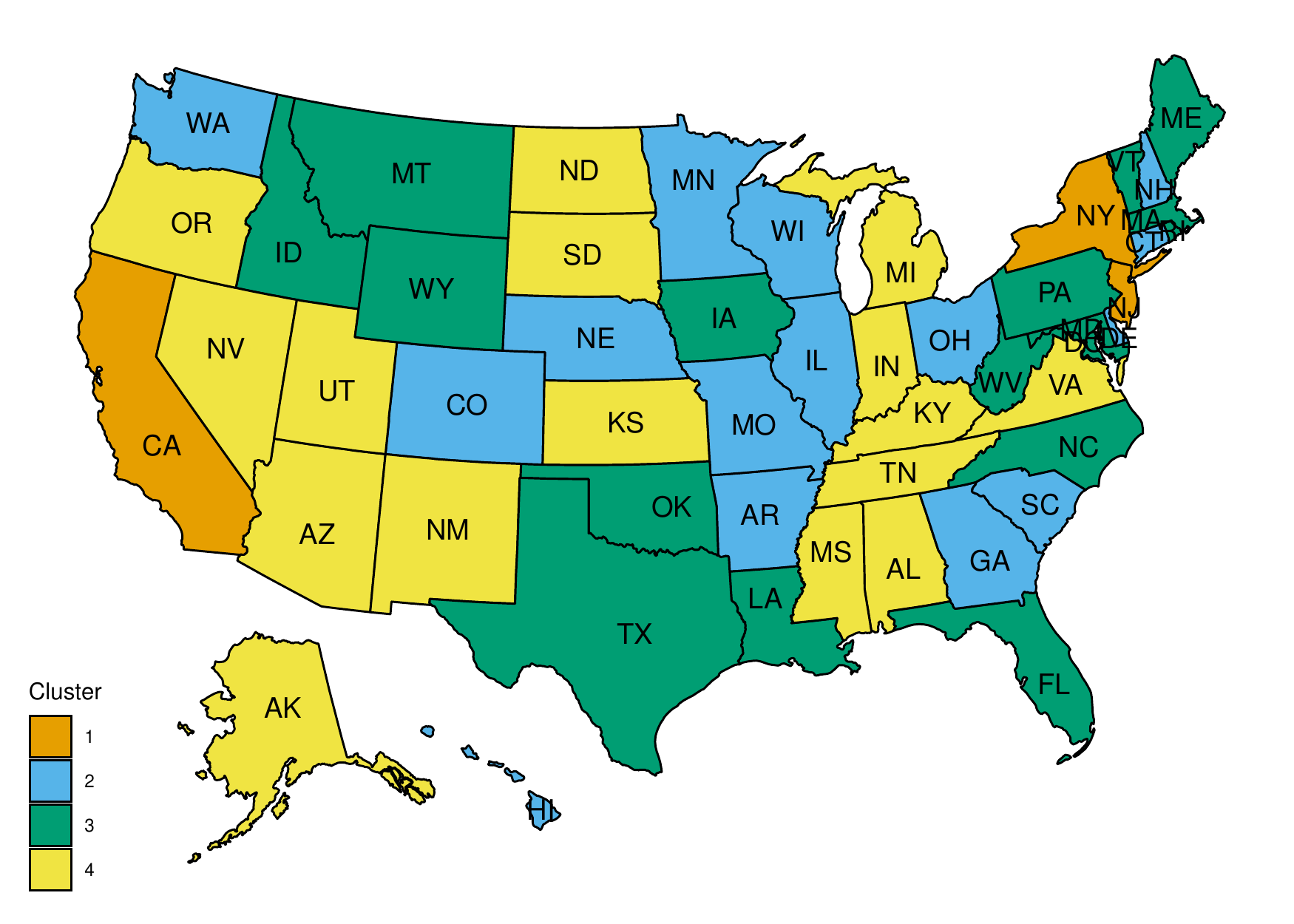}
  \caption{Illustration of the third partition setting with four
  clusters, where the clusters are composed of disjoint states
  and regions.}
  \label{fig:design3}
\end{figure}

We simulate the data using the spatial structure of the 51 states. 
\blue{We consider in total three partition settings with respective true number
of clusters~3, 5, and~4.}
The
first partition setting shown in Figure~\ref{fig:3clusterdesign} consists of
two
disjoint parts in the east and west coast. It is designed to mimic a rather
common economic pattern that geographically distant regions share similar ID
pattern, and geographical proximity is not the sole factor for determining
homogeneity in ID.  The second setting is the five-cluster partition in
Figure~\ref{fig:5clusterdesign}, where there are more true underlying clusters.
\blue{The third and final setting has four underlying clusters,
and the spatial constraint of clusters is ``weaker'' than those under the other
two designs,
composing of many spatially discontiguous states and regions,
as shown in Figure~\ref{fig:design3}.}

\begin{figure}[tbp]
    \centering
    \includegraphics[width = \textwidth]{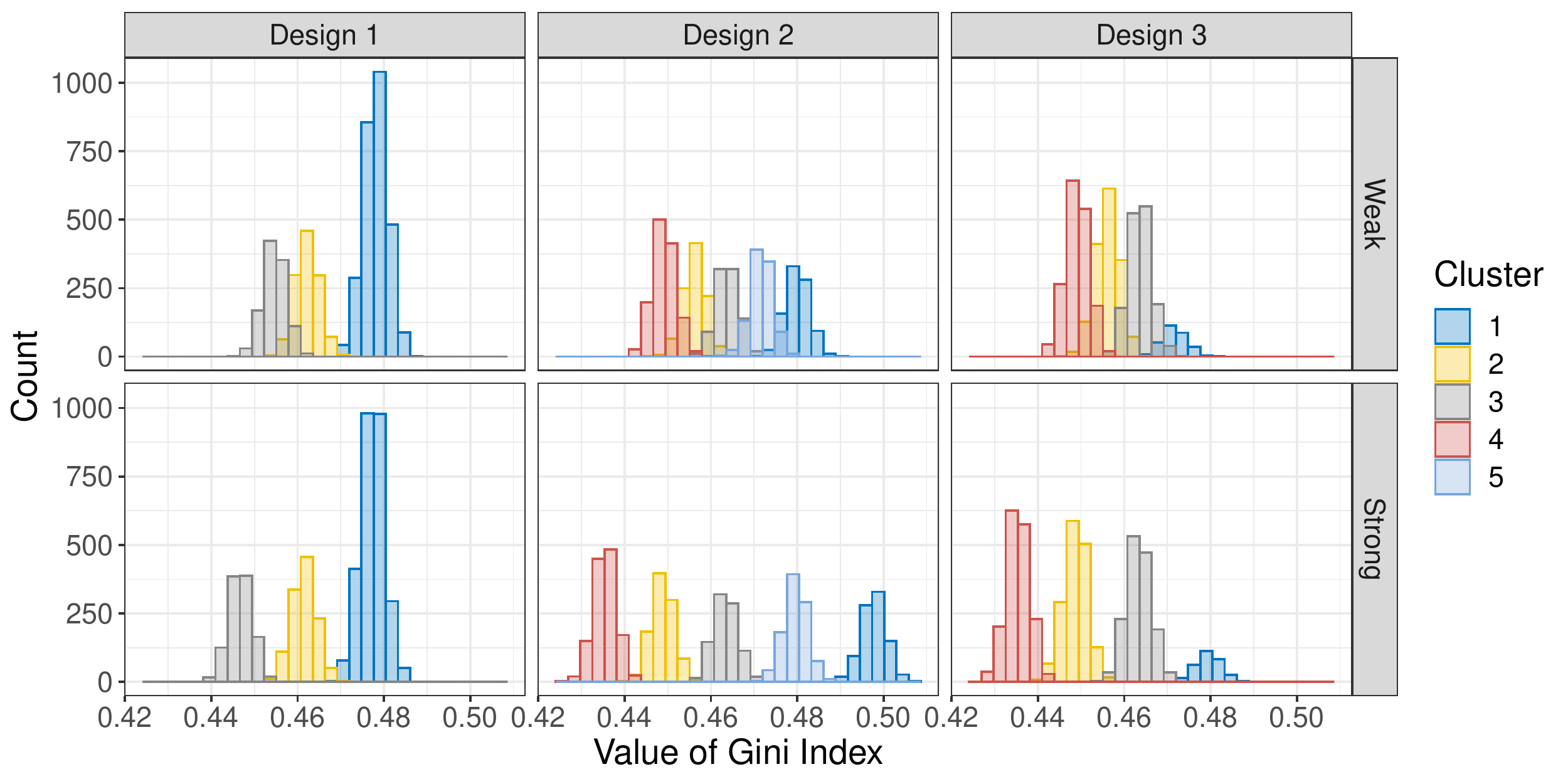}
\caption{Histograms of Gini indices calculated from the simulated state-wise
    income data (5,100 in each panel from 100 replicates) for weak and strong
    signals under the \blue{three} true partition settings.}
    \label{fig:ginihist}
\end{figure}

Following \cite{salem1974convenient}, we generate $10,000$ simulated
observations for each state from a Gamma distribution to mimic the long-tailed
pattern that is often observed in econometrics data. In addition, an additional
noise term following a Gamma distribution is added with probability so that the
ID within each state does not comprise a perfect fit to one certain
distribution. We assume each cluster has its own set of distribution parameters
shared by all states within it. The true values of the parameters are set so
that the Lorenz curves computed on the simulated data are highly similar to
those computed from real data (see Table~\ref{tab:simdesign}). We consider two
different parameter settings with small and large differences in income
distributions among clusters, corresponding to weak and strong signal designs,
respectively. For a total of 100 replicates, we show the Gini indices for
different clusters of both weak and strong signal designs in
Figure~\ref{fig:ginihist}, which clearly exhibits major and minor overlapping
among clusters, respectively.

The final clustering performance is evaluated using the estimated number of
clusters and ARI. The ARI is calculated using the final clustering result
selected by Dahl's method for each replicate, and we calculate an average ARI
over all replicates in each setting.  Computation of the ARI is facilitated
with
the \textsf{R} package \textbf{mclust} \citep{scrucca2016mclust}. In each
replicate of simulation, the outcome would be a clustering scheme of 51 states
into several clusters. If the number of unique clusters in the scheme for a
replicate equals the true number of clusters (3 in the first design, 5 in the
second), this replicate is counted towards one time that the number of clusters
is correctly inferred. We report the total counts of replicates with correctly
inferred number of clusters out of 100 replicates.

\begin{table}[tbp]
  \centering
  \caption{Simulation designs with weak and strong signals. The
  symbol~$\Gamma$ denotes Gamma distribution, and ``Bin'' denotes binomial
  distribution.}
  \label{tab:simdesign}
  \begin{tabular}{llccc}
  \toprule 
Design &Signal& Cluster  & Design\\ \midrule
    \textbf{Three Clusters} \\
     &     & 1  & 
              $\Gamma(1.15, 50000) + \TB(0.05)\cdot \Gamma(0.3, 50000)$ \\ 
   &   Weak & 2 & 
              $\Gamma(1.20, 50000) + \TB(0.05)\cdot \Gamma(0.3, 50000)$\\
    &    & 3 & $\Gamma(1.25, 50000) + \TB(0.05)\cdot \Gamma(0.3,
50000)$ \\ [0.5ex]
   &   & 1  & 
              $\Gamma(1.10, 50000) + \TB(0.05)\cdot \Gamma(0.5, 50000)$ \\ 
& Strong & 2&
              $\Gamma(1.20, 50000) + \TB(0.05)\cdot \Gamma(0.5, 50000)$\\
& & 3& $\Gamma(1.30, 50000) + \TB(0.05)\cdot \Gamma(0.5, 50000)$ \\ \midrule
    \textbf{Five Clusters} \\
     &     & 1  & 
              $\Gamma(1.10, 50000) + \TB(0.05)\cdot \Gamma(0.3, 50000)$ \\ 
   &    & 2 & 
              $\Gamma(1.15, 50000) + \TB(0.05)\cdot \Gamma(0.3, 50000)$\\
  & Weak     & 3 & $\Gamma(1.20, 50000) + \TB(0.05)\cdot \Gamma(0.3,
50000)$
 \\
    &    & 4 & 
              $\Gamma(1.25, 50000) + \TB(0.05)\cdot \Gamma(0.3, 50000)$\\
    &    & 5  & $\Gamma(1.30, 50000) + \TB(0.05)\cdot \Gamma(0.3,
50000)$\\ [0.5ex]
   &   & 1  & 
              $\Gamma(1.00, 50000) + \TB(0.05)\cdot \Gamma(0.5, 50000)$ \\ 
&  & 2&
              $\Gamma(1.10, 50000) + \TB(0.05)\cdot \Gamma(0.5, 50000)$\\
 & Strong & 3& $\Gamma(1.20, 50000) + \TB(0.05)\cdot \Gamma(0.5, 50000)$ \\
&  & 4&
              $\Gamma(1.30, 50000) + \TB(0.05)\cdot \Gamma(0.5, 50000)$\\
& & 5& $\Gamma(1.40, 50000) + \TB(0.05)\cdot \Gamma(0.5, 50000)$ \\ \midrule 
    \blue{\textbf{Four Clusters}} \\
   &    & 1 & 
              $\Gamma(1.15, 50000) + \TB(0.05)\cdot \Gamma(0.3, 50000)$\\
  & Weak     & 2 & $\Gamma(1.20, 50000) + \TB(0.05)\cdot \Gamma(0.3,
50000)$
 \\
    &    & 3 & 
              $\Gamma(1.25, 50000) + \TB(0.05)\cdot \Gamma(0.3, 50000)$\\
    &    & 4  & $\Gamma(1.30, 50000) + \TB(0.05)\cdot \Gamma(0.3,
50000)$\\ [0.5ex]
&  & 1&
              $\Gamma(1.10, 50000) + \TB(0.05)\cdot \Gamma(0.5, 50000)$\\
 & Strong & 2& $\Gamma(1.20, 50000) + \TB(0.05)\cdot \Gamma(0.5, 50000)$ \\
&  & 3&
              $\Gamma(1.30, 50000) + \TB(0.05)\cdot \Gamma(0.5, 50000)$\\
& & 4& $\Gamma(1.40, 50000) + \TB(0.05)\cdot \Gamma(0.5, 50000)$ \\
    \bottomrule
  \end{tabular}
  \end{table}

\subsection{Simulation Results} 

We first examine the inference results of the number of clusters, as well as
the
accuracy of clustering results from both MFM-fCluster and MRFC-MFM-fCluster.
Each parameter setting listed in Table~\ref{tab:simdesign} is run with 100
replicates. For MRFC-MFM-fCluster, $\lambda \in \{0.5, 1, 1.5, 2, 2.5, 3\}$ are
considered, and the best $\lambda$ value is selected using mDIC within each
replicate.
The graph distance \citep[GD;][]{bhattacharyya2014community} is used as the
distance measure to construct the neighborhood graph used in the Markov random
field model. Different upper limits of distance for two states to be considered
as ``neighbor'' are used for the two designs. For the first three-cluster
partition, the upper limit is set to~3. For the second five-cluster partition,
however, due to the relatively small true cluster sizes, an upper limit of~1,
i.e., only immediate neighbors, is adopted. \blue{Similarly, for the third
design, as the clusters have more disjoint components, and a state is more
likely to have neighbors that belong to a different cluster from its own,
we also consider an upper limit of~1.}

In addition to MFM-fCluster, we also consider two other competing methods.
In the first competing method, we treat the SRVFs derived from Lorenz curves as
vectors, and use $K$-means to cluster them. The second competing method is the
model-based clustering for sparsely sampled functional data proposed by
\cite{james2003clustering}, which is available in \textsf{R} package
\textbf{funcy}, and can be performed with function \textsf{funcit()} with
option
\textsf{method=``fitfclust''}. Clustering recovery performances of all three
methods are measured using ARI. For our proposed method, we present the average
of ARIs corresponding to the~$\lambda$ value selected by mDIC in each
replicate.
As neither $K$-means or model-based clustering
 can estimate the number of clusters but
instead require it to be provided, to make a fair comparison, we provide
the number of clusters inferred by each replicate corresponding to its
selected optimal~$\lambda$.

\begin{figure}[tbp]
    \centering
    \includegraphics[width = \textwidth]{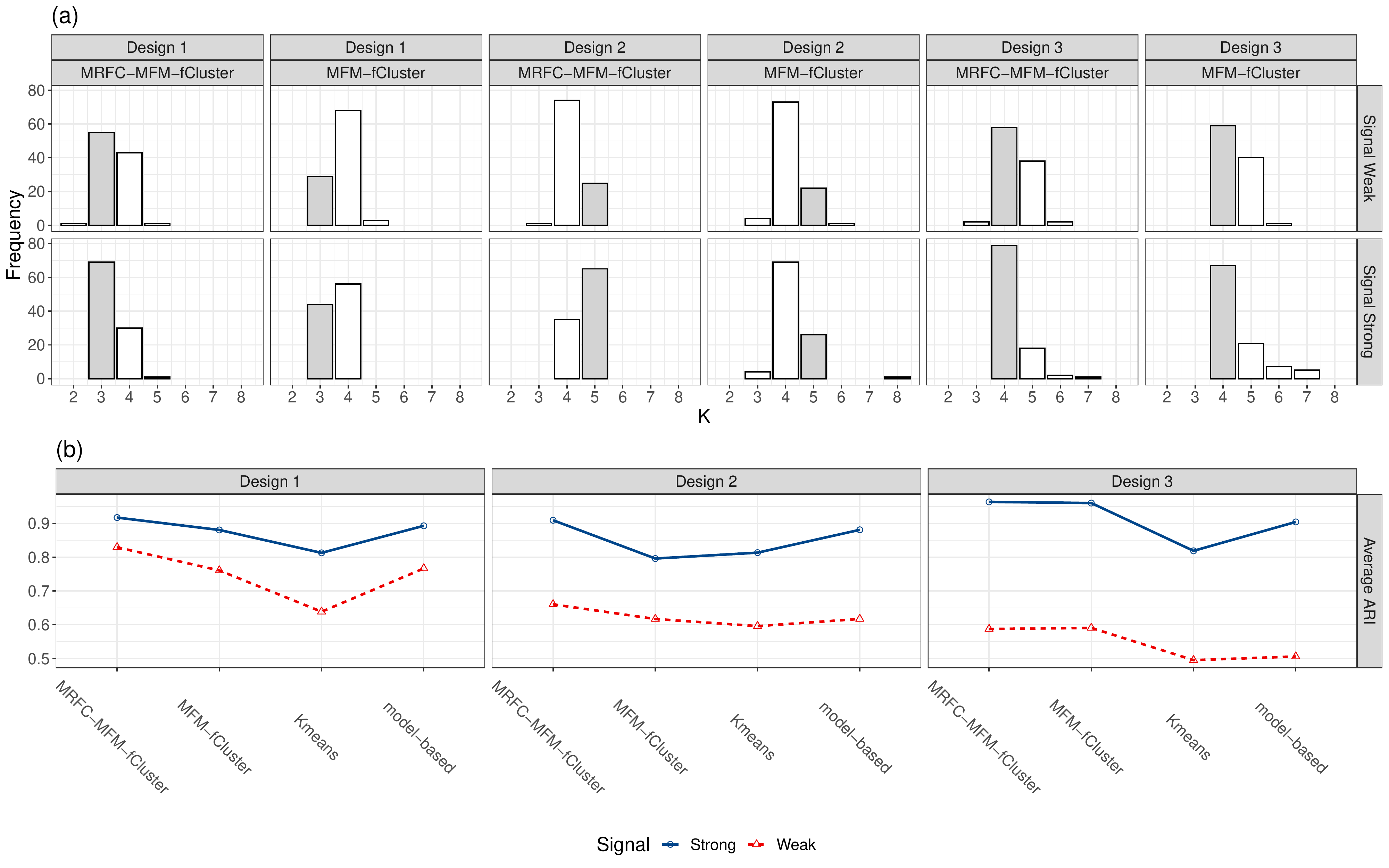}
    \caption{(a) Histogram of number of clusters inferred by MRFC-MFM-fCluster
    and MFM-fCluster under different designs and signal strength settings. The
    grey bars correspond to the correct number of clusters. (b) Plot of ARIs
for all four methods under different designs and signal strength settings.    }
    \label{fig:simu_comparison}
\end{figure}

Performances are visualized in Figure~\ref{fig:simu_comparison},
\blue{and the average optimal~$\lambda$'s selected by mDIC are presented in
Table~\ref{tab:lambda_bar}.} In
Figure~{\ref{fig:simu_comparison}(a)}, it can be seen that under design~1,
MFM-fCluster exhibits severe over-clustering, which produces four final
clusters
for more than~60 replicates under the weak signal setting, and more than~50
replicates in the strong signal setting. In contrast, even under the weak
signal
setting, MRFC-MFM-fCluster is able to correctly infer the true number of
clusters for more than~50 replicates, and a notable number of~69 for strong
signal. Under design~2, as cluster sizes are relatively small, it is rather
difficult for both MRFC-MFM-fCluster and MFM-fCluster to infer the number of
clusters under the weak signal setting, as can be seen from the
\blue{top middle} two
plots. With strong signal, however, MRFC-MFM-fCluster is able to correctly
identify the true number of clusters for more than half of simulation
replicates, while the performance of MFM-fCluster remains poor.
\blue{Under design~3, as the true clusters are ``messy'' in the sense that
there are no clear spatially contiguous states that belong to the same
cluster, the performance of MRFC-MFM-fCluster is similar to that of
MFM-fCluster in the weak signal setting. With strong signal, however,
MFM-fCluster again overclusters, producing for~67 replicates the correct~$K$,
while this number for MRFC-MFM-fCluster is~79.} In
Figure~{\ref{fig:simu_comparison}(b)}, our proposed method has the highest
average ARI over 100 replicates for all \blue{six} combinations of
signal and partition design. \blue{The model-based  functional clustering has
the second best performance in designs~1 and~2, and the third best performance
in
design~3, while MFM-fCluster has the second best performance in design~3,
and the third best in designs~1 and~2. In all cases, $K$-means performs the
worst.}

\blue{
\begin{table}[tbp]
    \centering
        \caption{Average~$\lambda$ selected by mDIC
        for~100 simulation replicates for each combination
        of singla strength and true cluster design.}
    \label{tab:lambda_bar}
    \begin{tabular}{lcccc}
    \toprule 
         & Design 1 & Design 2 & Design 3   \\ \midrule 
        Signal Weak & 1.535 & 1.715 & 1.630\\
        Signal Strong & 1.680 & 1.630  & 1.610\\
        \bottomrule
    \end{tabular}
\end{table}
}

In addition, computation times for all methods are benchmarked using \textsf{R}
package \textbf{microbenchmark} \citep{mersmann2019} on a desktop computer
running Windows 10 Enterprise, with i7-8700K CPU@3.70GHz using single-core
mode.
A total of 20 replicates are performed to compute the average running time for
each method. As expected, $k$-means takes the least time of 1.62 seconds due to
its simple iterative algorithm. Unlike the $k$-means which can only provide
clusters without making statistical inference of cluster memberships and sizes,
our proposed method utilizes conjugate forms for efficient Bayesian inference
that provides not only estimates of clusters but also their uncertainty
measures
at only a slightly higher computation cost. Indeed, it takes on average 20.79
seconds for one simulated dataset with 500~MCMC iterations, as in our empirical
studies 500 iterations are sufficient for the chain to converge and stabilize.
The model-based approach, however, takes more than three minutes to finish. Due
to the time-consuming nature of the model-based approach, the actual simulation
studies are conducted on a 16-core desktop computer using parallel computation.
The code is submitted for review and will be made publicly available at GitHub
after the acceptance of the manuscript.

\section{Analysis of PUMS Data}\label{sec:real_data}

In this section, we apply the proposed MRFC-MFM-fCluster to the analysis of US
households' income in 2017. Similar as in the simulation studies, the Lorenz
curves for all states are obtained for the functional clustering analysis.
Based
on \eqref{eq:srvf_def} and \eqref{eq:inner_product}, we get the inner product
matrix~$\bm{S}$. The spatial smoothing parameter~$\lambda$ is considered within
the range of \{0,0.2, 0.4,\ldots, 3\}, with $\lambda = 0$ corresponding to
MFM-fCluster. The upper limit for considering a state ``neighbor'' is
considered
within the range \{1, 2, 3\}. The mDIC is used to determine the optimal
combination of these two parameters. From the sensitivity analyses presented in
the supplemental material, $\gamma$ values are not particularly impactful on
the
clustering performance, and thus it is set to be consistent with in the
simulation studies described in Section~\ref{sec:mcmc_sampling}. We choose
following hyperparameters  $\alpha = 1$ and $\beta = 1$ which are consistent
with our simulation studies.

\begin{figure}[tbp]
\centering 
    \includegraphics[width = 0.8\textwidth]{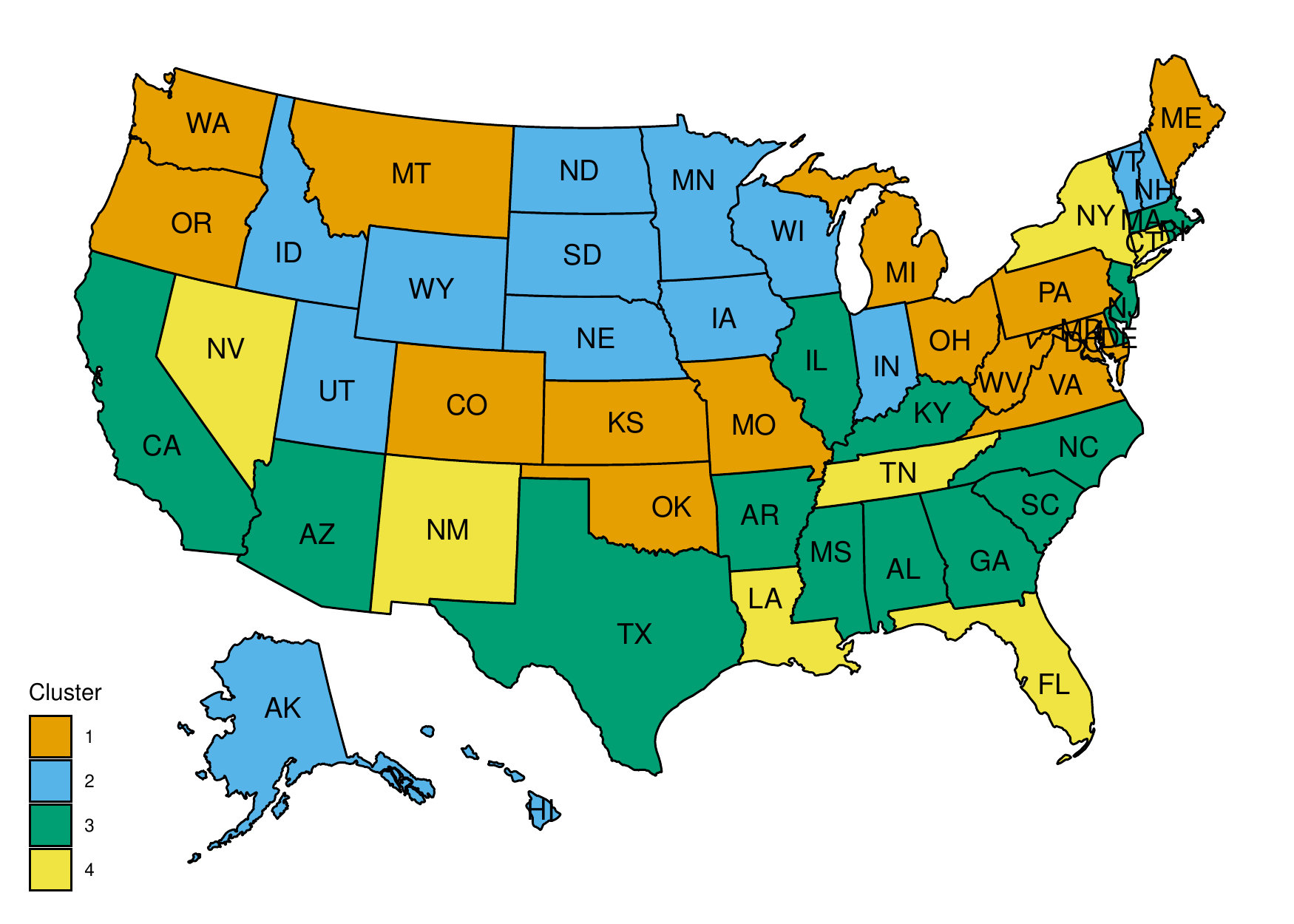}
\caption{Illustration of the four clusters identified by the proposed method
for the~51 states.}\label{fig:realdata}
\end{figure}

\begin{figure}[tbp]
\centering 
    \includegraphics[width = 0.6\textwidth]{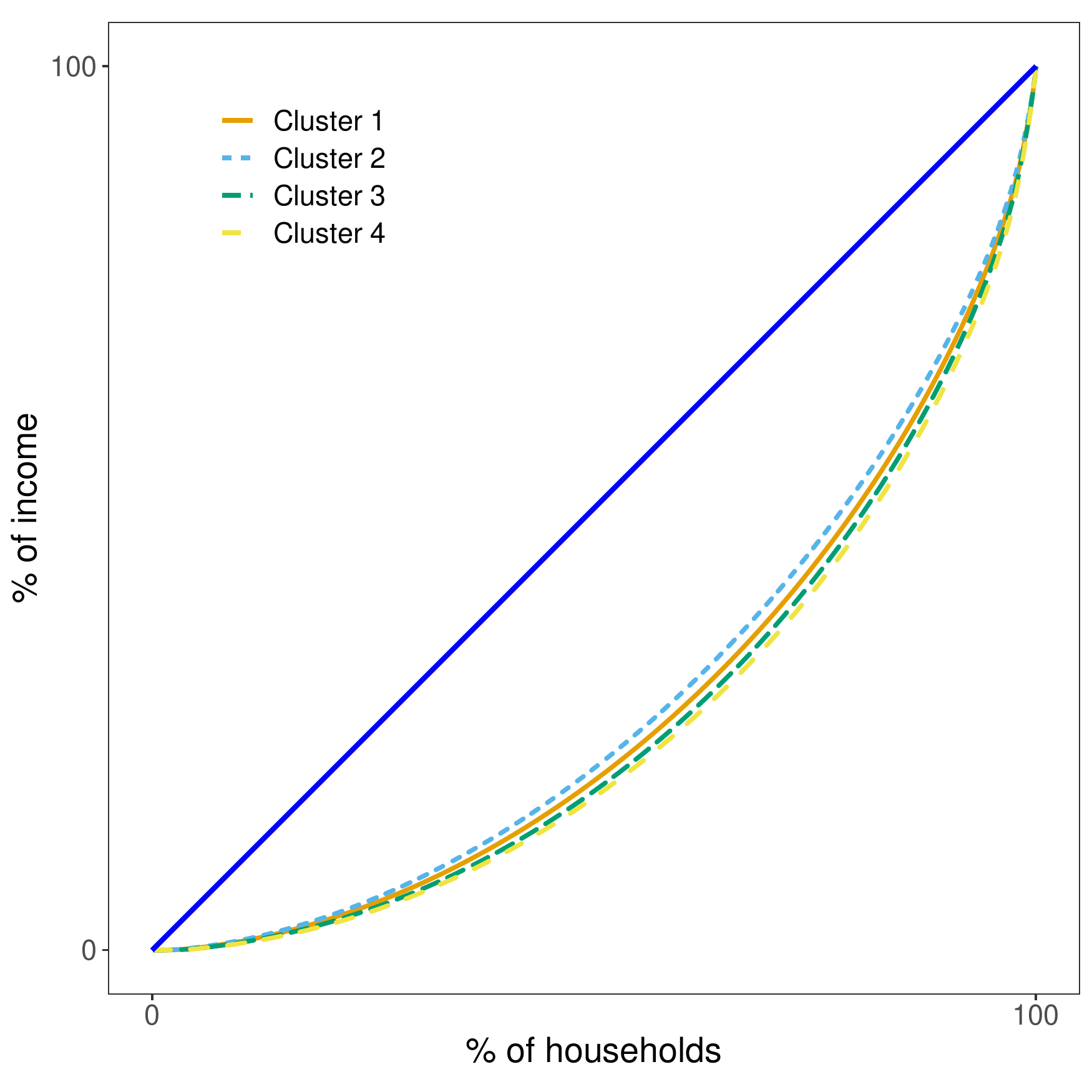}
\caption{Average Lorenz curves for states in the
four identified clusters.}\label{fig:lorenzbyCluster}
\end{figure}

The final model selected corresponding to the smallest mDIC value has $\lambda
=
0.8$ and upper limit~2 for defining neighbors. The final cluster configuration
is visualized in Figure~\ref{fig:realdata}. There are, respectively, 14, 14, 15
and 8 states in clusters 1, 2, 3 and~4. Cluster~4 tops in terms of income
inequality, and has an average Gini coefficient of 0.491. Cluster~2, with an
average Gini of 0.435, exhibits the most equal income distribution among the
four. Clusters~1 and~3 have average Gini values of 0.458 and 0.477.

One particularly important merit of our proposed method is that it allows for
globally discontinuous clusters. As shown in Figure~\ref{fig:realdata}, New
Mexico and Tennessee belong to the same cluster. Their 2017 Gini coefficients
are 0.4851 and 0.4858, respectively, which indicates these two states have very
similar IDs in terms of Gini coefficients. Based on 2010 American Community
Survey from U.S. Census Bureau \url{https://factfinder.census.gov/}, the Gini
coefficients of these two states have been historically very close. In addition,
there are several government policies that could be applied for different
clusters. For the states in Clusters~3 and~4, increasing the minimum wage and
expanding the earned income tax are two strategies for improving the equality of
ID. Most states in Cluster~1 have much lower median household income. Decreasing
the income tax will help increase their overall household income, which is at
the cost of minor sacrifice in ID equality. Furthermore, an increase in
government expenditures will help increase the household income directly for the
states in Cluster~1. For the states in Cluster~2, they have most balanced ID and
mid-level median household income. Most their government policies can be kept
for steady economy growth. From the clustering results, we still find that the
states with large metropolitan areas tend to have less balanced IDs, which is
consistent with the findings in \citet{glassmanincome}. \blue{Based on the
results shown in \citet{janikas2005spatial,rey2018bells} which analyzed the
income data collected from the Bureau of Economic Analysis, they claimed
that the states with high (low) levels of internal inequality tend to be located
next to other states with high (low) levels of inequality. This claim confirms
our spatial homogeneity patterns of IDs among different states. Taking Cluster 1
and Cluster 2 as examples, these two clusters have a large number of states
sharing
the same boundary, since they have low levels of internal inequality.}

The posterior estimate of $\bm{U}$ in~\eqref{eq:Func_MRFMFM} is
\begin{equation}
  \hat{U} = \begin{pmatrix}
  4.885 & 4.186 & 4.293 & 3.692 \\ 
  4.186 & 4.700 & 3.710 & 3.341 \\ 
  4.293 & 3.710 & 4.821 & 4.042 \\ 
  3.692 & 3.341 & 4.042 & 4.524 \\ 
  \end{pmatrix}.
  \label{eq:Umatrix}
\end{equation}
It is noticeable that the diagonal entries of $\bm{U}$ are larger than the
off-diagonal entries, which suggests the within-cluster similarity is much
higher than between-clusters similarities. Cluster~1 has least similarity with
Cluster~4 based on \eqref{eq:Umatrix}, which is consistent with the results
presented in Figure~\ref{fig:lorenzbyCluster}.

Finally, to make sure the cluster configuration presented here is not a random
occurrence but reflects the true pattern demonstrated by the data, we run 100
separate MCMC chains with different random seeds and initial values, and
obtained 100 final clustering schemes. The RI between each scheme and the
present clustering scheme in Figure~\ref{fig:realdata} is calculated, and they
average to~0.899, indicating high concordance of conclusion regardless of
random seeds. As suggested by a reviewer, we also use the
sequentially-allocated latent structure optimization (SALSO) algorithm
implemented in the \textsf{R} package \textbf{salso} \citep{dahl2020salso}
to check for undertainties in the presented clustering result. The details are
included in Section~4 of the supplemental material.

\section{Discussion}\label{sec:discuss}
In this paper, we proposed both mixture of finite mixtures (MFM) and Markov
random field constrained mixture of finite mixtures (MRFC-MFM) to capture
spatial homogeneity of ID based on functional inner product of
Lorenz curves.
Two efficient algorithms are proposed for the inference of the proposed
methods.
Parameter tuning is achieved using a modified version of DIC,
the popular Bayesian model selection criterion.
Comprehensive simulation studies are carried out to show MRFC-MFM achieves
better performance than the traditional MFM model in terms of spatial
homogeneity pursuit. It also outperforms the~$K$-means and model-based methods
under various designs, and the comparison of performance is relatively robust
under different choices of the spatial smoothing parameters. A case study using
the PUMS data reveals a number of important findings of IDs
across~51 states in US.

A few topics beyond the scope of this paper are worth further investigation. In
this paper, Fisher's $Z$-transformation of the inner product matrix is used.
Modeling the original inner product matrix is an interesting alternative in
future work.
\blue{Independence is assumed between elements
of the inner product matrix~$\bm{\mathscr{S}}$ for modeling and computation
simplicity and
convenience,
and extending SBM to incorporate such edge dependence similar to
\citet{yuan2018community} is an interesting but nontrivial problem devoted for
future research.}
In addition, tuning of $\lambda$ is criterion-based. Treating it as
an unknown parameter and proposing a prior in a hierarchical model for it may
improve the efficiency. Besides the geographical information, other auxiliary
covariates, such as demographic information, could also be taken into account
for clustering in our future work. \blue{While our clustering methods are based
on similarity matrix or dissimilarity matrix, the proposed MRFC-MFM clustering
prior model can be adapted to other hierarchical model settings, including the
case with multiple similarity matrices as responses
\citep{paul2016consistent,lei2020consistent}} Extending our prior on functional
data model
with basis coefficients \citep{suarez2016bayesian} is also \blue{another}
interesting
future
work.

\appendix
\section{Proof of Theorem \ref{thm:MRF-MFM}}\label{sec:proof}
Full conditionals $\Pi(\bm{\theta}_i\mid \bm{\theta}_{-i})$ from
\eqref{eq:MRFMFM1} can be obtained up to a constant as the product of the full
conditionals of each part
\begin{align}\label{eq:Func_MFM2}
  \Pi(\bm{\theta}_i\mid \bm{\theta}_{-i}) & \propto P(\bm{\theta}_i\mid
  \bm{\theta}_{-i})M(\bm{\theta}_i\mid \bm{\theta}_{-i}) \nonumber\\
  & \propto M(\bm{\theta}_i\mid \bm{\theta}_{-i})\sum_{k=1}^{K^*}(n_k^{(-i)} +
  \gamma)\delta_{\bm{\theta}^*_k} + M(\bm{\theta}_i\mid
  \bm{\theta}_{-i})\dfrac{V_n(K^*+1)}{V_n(K^*)}\gamma G_0(\bm{\theta}_{i})
  \nonumber\\
  & \propto M(\bm{\theta}_i\mid \bm{\theta}_{-i})\sum_{k=1}^{K^*}(n_k^{(-i)} +
  \gamma)\delta_{\bm{\theta}^*_k} + \dfrac{1}{Z_H}
  \dfrac{V_n(K^*+1)}{V_n(K^*)}\gamma G_0(\bm{\theta}_{i})\\
  & \propto \sum_{k=1}^{K^*}
  (n_k^{(-i)}+\gamma)\frac{1}{Z_H}\exp(-H(\bm{\theta}_i
  \mid\bm{\theta}_{-i}))\delta_{\bm{\theta}^*_k}
  (\bm{\theta}_{i}) + \dfrac{V_n(K^*+1)}{V_n(K^*)}
  \dfrac{\gamma}{Z_H}G_0(\bm{\theta}_{i})\nonumber
  \end{align}

As a direct characteristic from the defined cost function $H$ in
\eqref{eq:costuse}, the support of $H$ is the set of existing cluster
parameters
$\bm{\theta}^*_1,\ldots,\bm{\theta}^*_{K^*}$. When $\bm{\theta}_i$ is generated
from
the base distribution $G_0$, $H(\bm{\theta}_i\mid \bm{\theta}_{-i}) = 0$ and
$M(\bm{\theta}_i\mid \bm{\theta}_{-i}) = \dfrac{1}{Z_H}$. This results in the
derivation from the second to the third step above. 

The last step is simply to plug in $M(\bm{\theta}_i\mid \bm{\theta}_{-i})$ from
its definition.

\end{document}